\documentstyle[preprint,eqsecnum,aps]{revtex}
\begin{document}

%\documentstyle{article}
%\begin{document}
%\baselineskip 24pt

%\magnification=\magstep1
%\overfullrule=0pt
%\baselineskip=18pt

\draft
\title{Direct Scattering Processes  \\
and Signatures of Chaos in Quantum Waveguides}
\author{G. B. Akguc and L. E. Reichl}
\address{ Center for Studies in Statistical Mechanics and Complex Systems\\
The University of Texas at Austin\\
Austin, Texas 78712
}
\date{\today}
\maketitle
\begin{abstract}

The effect of direct processes on the statistical
properties of deterministic scattering processes in a chaotic 
waveguide is examined.
The single channel Poisson
kernel describes well the distribution of S-matrix eigenphases when
evaluated over an
energy interval. When direct processes are transformed away, the scattering
processes
exhibit universal random matrix behavior. The effect of chaos on scattering
wavefunctions,eigenphases, and time delays is discussed.

\end{abstract}
\narrowtext
\section{Introduction}
\label{sec:intro}

\bigskip
\bigskip
%-----------------------------------------------------------------------

In 1957, Wigner proposed the use of statistical measures to analyze
   complex nuclear scattering data \cite{kn:wigner}. It was soon found
that,  in some nuclear scattering data, the  spacing between
scattering resonances was distributed in a manner similar to the
spacing of energy
levels of a Gaussian random Hamiltonian \cite{kn:gary}. In the 1960's,
extensive work was done to develop a systematic theory of the statistical
properties of
random hermitian matrices
\cite{kn:brody.etal,kn:mehta} and random unitary matrices \cite{kn:4,kn:hua}.
The general criterion for
constructing these random matrices is that they minimize information.
In 1979, the appearance of random matrix-like behavior in quantum
systems was linked to
underlying chaos in
the classical deterministic dynamics of these  systems
\cite{kn:macdonald,kn:boh,kn:casati,kn:berry}.
     Since then a large
body of work
has developed linking the statistical properties of bounded and open
quantum systems to
underlying deterministic chaos \cite{kn:reichl,kn:guhr}.

Historically there have been two different approaches to describing the
statistics of
quantum scattering processes in chaotic systems. One approach
\cite{kn:weiden} begins
directly with the Hamiltonian and uses it to build the scattering
matrix. The other
approach
\cite{kn:mello3,kn:mello2}, begins directly with the scattering matrix. In both
cases, random matrices are used to describe scattering processes and the
statistical
properties of the eigenphases of the scattering matrix and partial delay
times can be
obtained and
compared to experiment.  There has been considerable success in recent
years to link
the results and predictions of these two approaches \cite{kn:brouwer}.

The connection between the statistical properties of scattering processes and
underlying chaos is not straightforward because scattering events may
involve either
reactive processes or direct processes.
Reactive scattering processes are those for which an incident
particle becomes engaged
with the dynamics in the reaction region, and may be
delayed there for a considerable time. Direct processes
are those for which the particle passes through the reaction region without
becoming significantly
engaged in the reaction region. One prediction of the random
matrix theory (RMT) of
scattering is that, if the scattering process is truly chaotic, the average
value of the S-matrix will be zero. However, if direct processes are
present this
will not happen.

In this paper, we will study the deterministic scattering of a 
quantum particle in
a two
dimensional ballistic waveguide which has a classically chaotic cavity
formed by a
ripple billiard connected to a single lead at one end (see Figure 1). The
ripple
billiard is particularly well suited to the use of the reaction 
matrix theory approach
to scattering \cite{kn:eisenbud}, 
because a
simple coordinate transformation \cite{kn:luna,kn:akg1} allows us to
construct a
Hamiltonian matrix and thus an
eigenvalue equation for the basis states inside the cavity.
We will compare the results of deterministic scattering from the chaotic
cavity to some
recent predictions of random matrix theory as regards the scattering
process. The open
geometry of the waveguide in Fig. (1) ensures that direct processes will play
an important role in scattering. We show that the contribution of the
direct processes
to the scattering can be transformed out and the statistical properties of
the reactive
part of the scattering process can be compared to random matrix 
predictions. Our
discussion focuses on quantum particles and we will use parameters 
appropriate for
electrons in waveguides made with GaAs, for which a number of 
experiments have been done
\cite{kn:marcus1,kn:chang,kn:bird}. However, our results also apply 
to electro-magnetic
waves in flat microwave cavities, because the eigenmodes in these 
cavities satisfy a
Schrodinger-like equation \cite{kn:graf,kn:stein,kn:stockmann}.

We begin in Section (2), by reviewing the  reaction matrix theory of
deterministic
scattering in the ballistic waveguide and we study some properties of the
cavity basis
states. In Section (3), we study various statistical properties of the
eigenphases of
the waveguide scattering matrix (S-matrix). We show that, when the cavity
dynamics is
chaotic, the deterministic S-matrix eigenphases exhibit level repulsion and
their
distribution is well described by a Poisson kernel.  In Section (4), we
compare the
distribution of partial delay times for the deterministic scattering
process to the
predictions of random matrix theory. 
Finally, in Section (5) we give
concluding
remarks.

\section{Reaction matrix description of scattering}

We will explore the behavior of a particle of mass, $m$, in a ballistic
waveguide as it scatters from the ripple cavity shown in Fig. (1). A 
particle with
energy,
$E$, enters the cavity from the left  along a straight lead which
has infinitely hard walls. The particle wave is reflected back to the left
by an
infinitely hard
wall located at $x=L$. The dynamics inside the cavity,
$0<x<L$, can range from mixed to fully chaotic as the ripple amplitude is
varied.
The Schrodinger equation, which describes propagation of a particle wave,
$\Psi(x,y,t)$, in the waveguide at time, $t$, is given by
\begin{equation}
i{\hbar}{{\partial}{\Psi}(x,y,t)\over {\partial}t}={\hat H}{\Psi}(x,y,t)=
{\biggl[}-{{\hbar}^2\over 2m}{\biggl(}{{\partial}^2\over
{\partial}x^2}+{{\partial}^2\over
{\partial}y^2}{\biggr)}+V(x,y){\biggr]}{\Psi}(x,y,t),
\end{equation}
where ${\hat H}$ is the total Hamiltonian of the particle in the waveguide
and ${\hbar}$
is Planck's constant. The potential,
$V(x,y)$, has the following
properties: $V(x,y)={\infty}$ for $(L{\leq}x<\infty)$;
$V(x,0)=\infty$ for $(-\infty{\leq}x{\leq}L)$;
$V(x,y=g(x))=\infty$ for $(0<x<L)$; and
$V(x,y=d)=\infty$ for $(-\infty<x<0)$;
where $g(x)=d+{\rm a}~{\sin}(5{\pi}x/L)$ gives the contour of the ripple, $d$ is
the
average width of the cavity, $L$ is the length, and a is the ripple
amplitude. In all subsequent sections, we take the particle mass to be the
effective mass
of an electron in GaAs, $m=0.067m_e$, where $m_e$ is the free electron mass.
An energy eigenstate, $|E{\rangle}$, with energy, $E$, satisfies
the equation, ${\hat H}|E{\rangle}=E|E{\rangle}$, and evolves as
${\Psi}(x,y,t)={\langle}x,y|E{\rangle}~{\rm e}^{-iEt/{\hbar}}$.

As shown in reference \cite{kn:akg1} the waveguide energy eigenstates can be
expressed in the form
\begin{equation}
{\langle}x,y|E{\rangle}={\sum_{j=1}^{\infty}}{\gamma}_j{\phi}_j(x,y)
+{\sum_{n=1}^{\infty}}{\Gamma}_n~{\Phi}_n(x,y).
\label{energystate}
\end{equation}
The states, ${\Phi}_n(x,y)$, are the basis states in the lead ($x{\leq}0$),
\begin{equation}
{\Phi}_n(x,y){\equiv}{\langle}x,y|{\Phi}_n{\rangle}=\sqrt{\frac{2}{d}}{\chi}_{n}
(x){\sin}{\biggl(}{n
\pi y\over d}{\biggr)}
\label{leadbasis}
\end{equation}
($n=1,2,...,{\infty}$). These will consist of both propagating and
evanescent modes, as
we will discuss below. The states,
${\phi}_j(x,y){\equiv}{\langle}x,y|{\phi}_j{\rangle}$ ($j=1,2,...,\infty$),
are the
basis states in the cavity ($0{\leq}x{\leq}L$). In practice, we truncate
the number of
cavity basis states to some large but finite number, $M$. The value of $M$
is determined
by the accuracy desired for the calculation.  The coefficients,
${\Gamma}_n$ and
${\gamma}_j$,  in Eq. (\ref{energystate}) are defined
      ${\Gamma}_n={\langle}{\Phi}_n|E{\rangle}$ and
${\gamma}_j={\langle}{\phi}_j|E{\rangle}$.

A complete set of basis states, ${\phi_j}(x,y)$, inside the ripple cavity has
been obtained by solving the Schrodinger equation in the cavity, using Neumann
boundary conditions (${\bigr(}{d{\phi}_j\over dx}{\bigl)}_{x=0}=0$)
at the entrance
($x=0$),
and Dirichlet boundary conditions, (${\phi}_j(x,y)=0$) at the walls. We
obtained the eigenfunctions, ${\phi_j}(x,y)$ and the associated eigenvalues,
       ${\lambda_j}$, using the transformation technique discussed in Ref.
\cite{kn:akg1},
Sect. V.

In Fig. (2.a), we show a Poincare surface of section  for
a classical particle in a closed ripple cavity with the same shape as 
in Fig. 1, and
with hard walls. We choose $d=101\AA$, $L=511\AA$, and ${\rm a}=1.0\AA$. We 
have plotted
Birkhoff coordinates, $p_x/p={\cos}(\alpha)$ versus $x$, each time
the particle  hits  the lower wall at point $x$ ($\alpha$ is the 
angle between the wall
and the momentum).  For these billiard dimensions, the classical 
phase space contains a
mixture of regular orbits, nonlinear resonances, and chaotic motion. 
If we increase the
ripple amplitude,
a, there is a range of values of
a for which the classical motion appears to become totally chaotic.  For the
scattering system (with one end of the cavity open) periodic orbits 
and nonlinear
resonance regions can't be reached classically by a particle that 
enters from the
left, but quantum mechanically tunneling into these regions is possible as
we will show Section IV (see also \cite{kn:her}).
In Figs.~(2.b-e), we show Husimi plots of quantum Poincare surfaces of section
\cite{kn:li} of  cavity basis states, ${\phi}_j(x,y)$, with eigenvalues
$\gamma_{983}=257.1934E_1$, $\gamma_{985}=257.9339E_1$,
$\gamma_{989}=258.6655E_1$, and $\gamma_{990}=258.9072E_1$, where
$E_1={{\hbar}^2{\pi}^2\over 2md^2}$. The  Husumi plots in Fig.~(2.c) and Fig.~(2.d) show
that these bases states reside on nonlinear resonances of the underlying classical phase
space. In Section IV, we will show that these bases states give the primary contribution
to sharp resonances in the transmission at these cavity parameters.

Inside the lead,  we must distinguish between propagating and
evanescent modes. The longitudinal component of the eigenstates in the
leads, for
propagating modes, can be written
\begin{equation}
{\Gamma}_n{\chi}_n(x)=\frac{a_n}{\sqrt{k_n}}
{\rm e}^{-ik_nx}+\frac{b_n}{\sqrt{k_n}}{\rm e}^{ik_nx},
\label{eq:chan1}
\end{equation}
where the wavevector, $k_n$ is given by $k_n=\sqrt{{2mE\over
{\hbar}^2}-{\bigl(}{n{\pi}\over d}{\bigr)}^2}$.
If there are $N$ propagating modes then $n=1,...,N$. Here we use a
unit current normalization. A particle propagating in the $n^{th}$ channel
has energy in
the interval, $n^2E_1{\leq}E{\leq}(n+1)^2E_1$, where
$E_1={{\hbar}^2{\pi}^2\over
2md^2}=0.0738$eV.
All remaining modes,
$n=N+1,...,\infty$,  are evanescent and can be represented in terms of  the
states,
\begin{equation}
{\Gamma}_n{\chi}_n(x)=\frac{c_n}{\sqrt{{\kappa}_n}}{\rm e}^{-{\kappa}_n|x|},
\label{eq:chan3}
\end{equation}
where ${\kappa}_n=\sqrt{{\bigl(}{n{\pi}\over d}{\bigr)}^2-{2mE\over
{\hbar}^2}}$.
In Ref. \cite{kn:akg1},
we showed
that the evanescent modes for this system do not make a significant
contribution to the
scattering properties that we are interested in. Therefore we will neglect the
evanescent modes in subsequent sections.

As shown in Ref. \cite{kn:akg1}, The eigenvalue equation,
${\hat H}|E{\rangle}=E|E{\rangle}$, leads to the relation,
\begin{equation}
{\gamma}_j={{\hbar}^2\over 2m}{1\over
(E-{\lambda}_j)} {\sum_{n=1}^N}~{\phi}^{*}_{j,n}(0){\biggl(}{d{\chi}_n\over
dx}{\biggr)}_{x=0}{\Gamma}_{n}.
\label{eq:reac1}
\end{equation}
Also, continuity of energy eigenstates at $x=0$ gives
\begin{equation}
{\Gamma}_n{\chi}_n(0)={\sum_{j=1}^M}{\gamma}_j{\phi}_{j,n}(0)=
{\sum_{n'=1}^N}~{R}_{n,n'}~{\biggl(}{d{\chi}_n\over
dx}{\biggr)}_{x=0}{\Gamma}_{n'}.
\label{eq:reac2}
\end{equation}
where
\begin{equation}
{R}_{n,n'}={{\hbar}^2\over 2m}{\sum_{j=1}^M}
{{\phi}_{j,n'}^*(0){\phi}_{j,n}(0)\over (E-{\lambda}_j)}
\label{eq:reac3}
\end{equation}
is the $(n,n')^{th}$ matrix element of the reaction matrix  \cite{kn:eisenbud}.
The quantity, ${\phi}_{j,n}(0)$ is a measure of the overlap between
the  $j^{th}$
cavity state, and the $n^{th}$ channel in the lead, evaluated at the interface,
\begin{equation}
{\phi}_{j,n}(0)=\sqrt{2\over
d}{\int_{0}^d}~dy~{\phi}_{j}(0,y)~{\sin}{\bigl(}{n{\pi}y\over d}{\bigr)}.
\label{eq:reac4}
\end{equation}

Let us now form an $N{\times}1$ column matrix, ${\bar b}$ (${\bar a}$)
whose matrix
elements consist of the $N$ probability amplitudes ${\{}b_n{\}}$
(${\{}a_n{\}}$) of the
outgoing (incoming) {\it propagating} modes.   The waveguide scattering matrix
(S-matrix), ${\bar S}$, is a $N{\times}N$ matrix which connects the
incoming propagating
modes to the outgoing propagating modes, ${\bar b}={\bar S}{\cdot}{\bar a}$.

For the case when there are $N$ propagating modes in the lead we can obtain an
$N{\times}N$ S-matrix which may be written
\begin{equation}
{\bar S}=-{({\bar 1}_N-i{\bar {\rm K}})
\over ({\bar 1}_N+i{\bar {\rm K}})},
\label{eq:scat1}
\end{equation}
where ${\bar 1}_N$ is $N{\times}N$ unit matrix, and the $N{\times}N$
matrix, ${\bar {\rm
K}}$, has matrix elements,  ${\bar {\rm
K}}_{n,n'}=\sqrt{k_n}R_{n,n'}\sqrt{k_{n'}}$ and can be written
\begin{equation}
{\bar {\rm K}}={\bar w}^\dagger{\cdot} \frac{1}{E{\bar 1}_M-{\bar
H}_{in}}{\cdot}{\bar w}.
\label{eq:scat1.a}
\end{equation}
In Eq. (\ref{eq:scat1.a}),  ${\bar 1}_M$ is the $M{\times}M$ unit matrix,
${\bar H}_{in}$
is an $M{\times}M$ diagonal matrix  formed with the eigenvalues,
$\lambda_j$ ($j=1,...M$)
in the cavity, and ${\bar w}$ is an $M{\times}N$ coupling matrix,
\begin{equation}
{\bar w}\equiv \pmatrix{w_{1,1}& \ldots& w_{1,{N}}\cr
\vdots&&\vdots \cr w_{M,1}& \ldots& w_{M,N}\cr },
\label{eq:scat2}
\end{equation}
where $w_{j,n'}=\phi_{j,n'}(0)\sqrt{k_{n'}}$.   With some algebra, the
S-matrix can also
be written  in the form
\begin{eqnarray}
{\bar S}=-{\biggl(}{\bar 1}_N-2i{\bar w}^\dagger {\cdot}\frac{1}{E{\bar
1}_M-{\bar
H}_{in}+i{\bar w}{\cdot}{\bar
w}^\dagger}{\cdot}{\bar
w}{\biggr)}.
\label{eq:scat3}
\end{eqnarray}
In Ref. \cite{kn:akg1}, we showed that, if evanescent modes are included, an
additional term appears in the denominator in Eq. (\ref{eq:scat3}).

The reaction matrix approach to waveguide scattering
provides a very efficient means of computing the statistical properties of the
scattering process because the Schrodinger equation only needs to be solved
once to
obtain the basis states and eigenvalues in the cavity. Using these values,
the S-matrix
can then be obtained at all other particle energies, $E$. Typically for the
ripple cavity in
Figure (1), we can obtain the scattering matrix at $10^5$ different values
of incident
energy in a reasonable amount of time on a Cray machine.

One of the goals of this paper is to compare the statistical properties of the
deterministic scattering process in the ripple cavity to statistical
properties of a
hypothetical scattering process in which ${\hat H}_{in}$ is replaced by a diagonal
matrix, ${\hat H}_{goe}$, composed of the $M$ eigenvalues of an
$M{\times}M$
Gaussian Orthogonal Ensemble (GOE) Hamiltonian, ${\hat H}_{goe}'$, and the
$N$ columns of
the coupling matrix, ${\bar w}$, are replaced by $N$ of the $M$ normalized
eigenvectors
of ${\hat H}_{goe}'$ to yield a coupling matrix, ${\hat w}_{goe}$
\cite{kn:brody,kn:mehta}. In this random matrix theory approach, the strength of the
coupling between the cavity and the lead is given by the phenomenological
parameter,
$g$. The parameter, $g$, does not appear in the deterministic scattering process. For
deterministic scattering the strength of the coupling is entirely determined by the
geometry and the potentials at the interface. The  scattering matrix, obtain from RMT,  can
then be written
\begin{eqnarray}
{\bar S}_{goe}=-{\biggl(}{\bar 1}_N-2ig{\bar w}_{goe}^\dagger
{\cdot}\frac{1}{E{\bar
1}_M-{\bar H}_{goe}+ig{\bar w}_{goe}{\cdot}{\bar
w}_{goe}^\dagger}{\cdot}{\bar
w}_{goe}{\biggr)}.
\label{eq:scatgoe}
\end{eqnarray}
It was shown in Ref. (\cite{kn:vaarbar}), using supersymmetry
techniques, that for
the case when the distribution of energy eigenvalues of ${\hat H}_{goe}$
is centered at $E=0$ and
$M{\rightarrow}\infty$, the average S-matrix can be written
\begin{equation}
{\langle}{\bar S}_{goe}{\rangle}=s{\bar 1}_N
~~~{\rm with}~~~s=\frac{1-g[iE/2+\pi\nu(E)]}
                    {1+g[iE/2+\pi\nu(E)]},
\label{eq:avsmat2}
\end{equation}
where $\nu(E)={\pi}^{-1}\sqrt{1-(E/2)^2}$ is the average density of energy
eigenstates. It is useful to introduce the quantity
\begin{equation}
\mu={\mu}_r+i{\mu}_i={1-s\over 1+s^*}=g{\nu}{\pi}+ig{E\over 2},
\label{eq:avsmat3}
\end{equation}
where ${\mu}_r=g{\nu}{\pi}$ and ${\mu}_i=g{E\over 2}$,
respectively, are
the real and imaginary parts of $\mu$.  The case when
$g=1$, corresponds to ideal coupling. In the neighborhood of
$E{\approx}0$, the
eigenvalues of ${\hat H}_{goe}$ have a constant density, ${1\over 2\pi}$,
and the
average S-matrix ${\langle}{\bar S}_{goe}{\rangle}=0$. When
$g{\neq}1$, the
average S-matrix cannot be zero.

\section{Eigenphases of the Scattering Matrix }

We have analyzed some of the statistical properties of the eigenphases of
the S-matrix
for the case of  deterministic  scattering from the ripple cavity for the
cases when the
internal dynamics in the cavity is completely chaotic and when it is
near-integrable. In
this section, we consider the
energy interval  $256E_1{\leq}E{\leq}289E_1$ when 16 channels are present
in the lead.
The S-matrix is a $16{\times}16$ matrix, and for each value of incident
energy it has 16
eigenvalues, ${\rm e}^{i{\delta}_{\alpha}}$ (${\alpha}=1,...,16$) and 16
eigenfunctions, $|{\delta}_{\alpha}{\rangle}$ (${\bar
S}|{\delta}_{\alpha}{\rangle}=
{\rm e}^{i{\delta}_{\alpha}}|{\delta}_{\alpha}{\rangle}$). The S-matrix is
unitary so
the eigenfunctions, $|{\delta}_{\alpha}{\rangle}$, form a complete
orthonormal set.
We can use the orthonormality of the eigenfunctions to follow each eigenphase,
${\delta}_{\alpha}$, continuously as a function of energy
\cite{kn:akg}. The eigenfunctions, for two S-matrices evaluated at nearby
energies, will
be approximately orthogonal if they do not belong to the same eigenphase.
Thus we
can plot each of the 16 different eigenphases as a function of the 
incident energy.
These are
shown in Figure (3) where
the eigenphases, which are defined mod $2\pi$, are ``unwrapped" and allowed
to evolve
continuously as a function of energy. In Fig. (3.a), we show the case with
ripple
amplitude, ${\rm a}=25\AA$, where the classical cavity dynamics is chaotic, and in Fig.
(3.b) we show the
case ${\rm a}=1\AA$ where the classical cavity dynamics is mixed (see Fig. (2)). The
case of mixed dynamics shows many more
abrupt changes of phase as a function of energy than the chaotic case. This is due to
the fact that the mixed dynamics has many long lived resonances not found in the chaotic
case. This was also seen in Ref. \cite{kn:akg,kn:her}. We shall return to 
this feature in
Section (IV).

Below we first discuss the effect of direct processes on the distribution of
eigenphases, and  then we determine the distribution of nearest 
neighbor spacings of
these eigenphases.

\subsection{Distribution of Eigenphases }

      When a scattering process has a non-zero average S-matrix, ${\langle}\bar
S{\rangle}$, it  indicates that direct processes may play a 
significant role in the
scattering process. Direct processes are generally scattering events which
do not interact significantly with the reaction region (cavity) 
\cite{kn:mello4}. When
direct processes are present, the distribution of S-matrix elements 
that minimizes
information about the scattering process is the Poisson kernel. For 
the case of an $N$
channel process whose dynamics is time reversal invariant, the 
Poisson kernel has the
form,
\begin{equation}
P_N({\bar S})=\frac{1}{{\Omega}} \frac{[{\rm
Det}(1-{\langle}S{\rangle}^{*}{\langle}S{\rangle}]^{(N+1)/2}} {|{\rm
Det}(1-{\langle}S{\rangle}^{*}{\bar S})|^{(N+1)}},
\label{eq:poiskern1}
\end{equation}
where ${\Omega}$ is a normalization factor that ensures that the Poisson kernel
satisfies the normalization condition, ${\int}d{\bar S}P_N({\bar S})=1$.

The S-matrix  can be diagonalized by a unitary matrix, ${\bar U}$,
and,  as mentioned
earlier, the eigenvalues
of the S-matrix are denoted, ${\rm e}^{i{\delta}_{\alpha}}$,
${\alpha}=1,...,N$.  In
terms of the eigenphases, ${\delta}_{\alpha}$, the normalization condition
for the
Poisson kernel, Eq. (\ref{eq:poiskern1}), can be written
\begin{eqnarray}
{\int}d{\bar S}P_N({\bar S})={\int}...{\int}d{\delta}_1...d{\delta}_N~
P_N({\delta}_1,...,{\delta}_N)\nonumber\\
=\frac{1}{{\Omega}_U}{\int}...{\int}d{\delta}_1...d{\delta}_N~
{\times}{\prod_{1{\leq}{\alpha}<{\alpha}'{\leq}N}}|{\rm
e}^{i{\delta}_{\alpha}}-{\rm
e}^{i{\delta}_{\alpha}'}|\nonumber\\
{\times}{\biggl(}{(1-s^*s)^N\over
{\prod_{{\alpha}=1}^N}(1-s^*{\rm e}^{i{\delta}_{\alpha}}) (1-s{\rm
e}^{-i{\delta}_{{\alpha}'}})}{\biggr)}^{(N+1)/2}=1,~~
\label{poiskern2}
\end{eqnarray}
where ${\Omega}_U$ is a normalization constant.
Note that $P_N({\delta}_1,...,{\delta}_N)$ is the joint probability density
to find the
angles, ${\delta}_{\alpha}$, in the intervals
${\delta}_{\alpha}{\rightarrow}{\delta}_{\alpha}+d{\delta}_{\alpha}$,
(${\alpha}=1,...,N$).

In Ref. \cite{kn:fyo2}, it is shown   that if the following change of angles is
introduced
\begin{equation}
{\tan}(\frac{{\theta}_{\alpha}}{2})={1\over
g{\pi}{\nu}}{\biggl(}{\tan}(\frac{{\delta}_{\alpha}}{2})+g{E\over
2}{\biggr)},
\label{eq:avsmat}
\end{equation}
and if one assumes ideal coupling, $g=1$, then Eq. (\ref{poiskern2})
reduces to
\begin{eqnarray}
{\int}d{\bar S}P_N({\bar
S})={\int}...{\int}d{\theta}_1...d{\theta}_N~P_N({\theta}_1,...,{\theta}_N)~~~~~~~
~~\nonumber\\
=\frac{1}{{\Omega}_U}{\int}...{\int}d{\theta}_1...d{\theta}_N~
{\prod_{1{\leq}{\alpha}<{\alpha}'{\leq}N}}|{\rm e}^{i{\theta}_{\alpha}}-{\rm
e}^{i{\theta}_{{\alpha}'}}|,
\label{poiskern3}
\end{eqnarray}
which is just the distribution for the Circular Orthogonal Ensemble (COE)
 \cite{kn:4}, \cite{kn:hua}.  Thus, even for scattering processes which
include direct
processes, it is possible in principle to transform away the direct
processes and compare
the eigenphase distribution with that of COE (note that similar
ideas first appeared in literature in Ref.~\cite{kn:mello2}). It is important to note that
the transformation that removes direct processes is not the same as the unfolding process
that occurs on the energy spectrum of bounded systems to give a constant average density. 

In our subsequent analysis, the case of scattering with only a single
channel will be  useful for analyzing data. For single channel scattering ($N=1$),  the
S-matrix reduces to the complex
function, $S={\rm e}^{i\delta}$, and  the Poisson kernel, reduces to \cite{kn:mello3}
\begin{equation}
P_1{s}(\delta)=\frac{1}{2\pi} \frac{[(1-s^{*}s]}
{|(1-s^{*}{\rm e}^{i\delta})|^{2}},
\label{eq:poiskern4}
\end{equation}
with normalization condition, ${\int_{-\pi}^{\pi}}d{\delta}P_1(\delta)=1$.
Under the
transformation above, $P_1(\delta){\rightarrow}P_1(\theta)=1/2\pi$,
which is the COE prediction.

Having obtained numerical values of the eigenvalues,
${\rm e}^{i{\delta}_{\alpha}}$, as a function of energy,  we can compute
an average
value for each of the  16 eigenvalues,
\begin{equation}
s_{\alpha}={\langle}{\rm e}^{i{\delta}_{\alpha}}{\rangle}={1\over
\eta}{\sum_{k=1}^{\eta}}{\rm
e}^{i{\delta}_{\alpha}(E_k)},
\label{eq:poiskern5}
\end{equation}
where ${\eta}$ is the number of energy values used. The apparently continuous
eigenphase curves actually consist of about
$40,000$ discrete energy points. The approximate orthonormality of S-matrix 
eigenvectors for neighboring energies has been used to sort the eigenphases. 
Thus, the eigenphases and eigenvectors have an energy interval over which they are
correlated, and we have used that fact in our sorting process. On the other hand, this 
correlation of the S-matrices at neighboring energies can prevent us from obtaining
statistics that can be compared to RMT predictions. Comparison to RMT requires
use of independent data points. Therefore, in order to study the statistical properties
of the eigenphases, we must choose values of the eigenphases separated in energy a
distance greater than the correlation length. For each eigenphase curve we select points
which have an energy spacing, 
$\Delta E=0.495E_1$. We choose this spacing based on an analysis of the delay time
correlation discussed in Section IV. (The delay time auto-correlation function is the second
derivative of the eigenphase auto-correlation function.) 

      We have computed a histogram of the number of eigenphases,
$N(\delta)={\eta}P_N(\delta)$, versus value of eigenphase,
${\delta}_{\alpha}$, in the 16 channel region, where ${\eta}$ is the number of data
points. In order to improve the statistics, we use data from four different ripple
amplitudes, 
${\rm a}=25\AA, ~30\AA,~35\AA,45\AA$, all of which lie in the chaotic regime. All
eigenphases lie in the energy interval, $256<E/E_1<289$ and have energy spacing, $\Delta
E=0.495E_1$. Thus the histogram includes $67 X 16 X 4=4288$ data points.
We have found that the distribution of eigenphases, along a given eigenphase curve, is well
described by the Poisson kernel for the single channel case.  We proceed as follows. We 
compute the average eigenvalue, $s_{\alpha}={\langle}{\rm
e}^{i{\delta}_{\alpha}}{\rangle}$, for each eigenphase curve.   
We can form a histogram using the 67 data points from a single eigenphase curve. We do
this for each of the 64 eigenphase curves, using the same number of bins and bin width
for each curve. We then add these 64 histograms together to form a single histogram,
which is   shown in Fig. (4.a).  The solid line in Fig.
(4.a), is the single channel Poisson kernel, $P_1(\delta)$, but with
${\langle}s{\rangle}={1\over 64}{\sum_{\alpha=1}^{64}}s_{\alpha}$.
We can use Eq. (\ref{eq:avsmat}) to
transform away the effects of the direct interactions
and find the
distribution of the transformed angles, ${\theta}_{\alpha}$. If we use Eq.
(\ref{eq:avsmat3}) and
obtain ${\mu}$ from the numerically calculated values of
$s_{\alpha}={\langle}{\rm
e}^{i{\delta}_{\alpha}}{\rangle}$, then  the transformed angles,
${\theta}_{\alpha}$,
are given by
\begin{equation}
{\tan}(\frac{{\theta}_{\alpha}}{2})={1\over
{\mu}_r}{\biggl(}{\tan}(\frac{{\delta}_{\alpha}}{2})+{\mu}_i{\biggr)},
\label{eq:avsmat4}
\end{equation}
with ${\mu}_r$ and ${\mu}_i$ computed numerically from $s_{\alpha}$.
In Fig.~(4.b), we show how the histogram in Fig. (4.a) changes if we
transform each
eigenphase, ${\delta}_{\alpha}$, using Eq. (\ref{eq:avsmat4}). In this case
distribution
is approximately constant and equal to the area,
$\frac{N}{2\pi}$.  Thus, having transformed away the contribution
from the direct
processes, we obtain the COE eigenphase distribution for this chaotic
scattering
process, with fairly high confidence level.

For ${\rm a}=1\AA$ the plots of the eigenphase distributions look very similar
to the chaotic case shown in Fig. (4), and it appears that the
eigenphase distribution is not as sensitive an indicator for
underlying chaos as is
the nearest
neighbor spacing distribution, at least with this type of analysis.

\subsection{Nearest Neighbor Eigenphase Spacing }

In this section we consider the nearest neighbor spacings between eigenphases,
${\delta}_{\alpha}$, of the
scattering matrix
for the 16 channel case in the energy interval $256<E/E_1<289$. For any
given value of the energy, the S-matrix only has 16 eigenphases.
However, we can form a
histogram of
nearest neighbor eigenphase spacings if we obtain eigenphase spacings for a
sequence of different energies in the range,
$256<E/E_1<289$.
In Fig. (5.a) we show histograms of 1005 (15X67)
nearest neighbor spacings for eigenphases
computed at energy increments,
${\Delta}E=0.495E_1$ and obtained by averaging over histograms
 for each ripple amplitude, ${\rm a}=25\AA,~30\AA,~35\AA,~45\AA$.
We have fit the histogram to the Brody distribution,
\cite{kn:brody}
\begin{equation}
P_B({\sigma})=A{\biggl(}{{\sigma}\over
{\langle}{\sigma}{\rangle}}{\biggr)}^{\beta}\exp{\biggl[}-{\xi}
{\biggl(}{{\sigma}\over {\langle}{\sigma}{\rangle}}{\biggr)}^{1+\beta}{\biggr]}
~~~{\rm with}~~~
\xi={\biggl[}{1\over
{\langle}{\sigma}{\rangle}}\Gamma{\biggl(}\frac{2+\beta}{1+\beta}{\biggr)}{\biggr]}^{1+\beta},
\end{equation}
${\langle}{\sigma}{\rangle}$ is the average spacing between nearest neighbor
eigenphases, and ${\Gamma}(x)$ is the Gamma function. In Fig. (5.a), the solid line
is a fit to the Brody distribution for $\beta=0.635$.
In Fig.~(5.b) same calculation is performed after direct processes were
transformed away. In this case $\beta=0.865$.
Note that the GOE prediction ($\beta=0.95$) for the closed system eigenvalue
spacings is fairly close to our value of $\beta$, after the effects of 
direct processes are transformed  away. In Ref. \cite{kn:li}, the nearest neighbor
energy  eigenvalue spacings for a closed ripple billiard were fit to the Brody
distribution with  $\beta=0.806$. In that case, the deviation from GOE predictions was
found to be due to  bouncing ball orbits. Our result also contains bouncing ball
contributions.  It is  useful to note that in our scattering system there is no long
range energy correlation  for 16 channel region (we have explicitly removed energy
correlations by taking data points  at large energy increments), in contrast to the case
reported in reference \cite{kn:dietz}

We also obtained a nearest neighbor spacing histogram  for the case with mixed
phase space in the 16 channel energy interval.  In
Fig.~(6.a) we show the histogram of nearest neighbor eigenphase spacings for
${\rm a}=0.5\AA,~ 1\AA,~ 2\AA,~ 3\AA,~ 4\AA,~ 5\AA$ in the
energy interval $256<E/E_1<289$ before direct process are transformed away.
We use an energy spacing ${\Delta}E=0.495E_1$ and obtain 1005 data points for 
each amplitude. We then average over the histograms  for the six ripple amplitudes.
The solid line is a fit to the Brody
distribution for ${\beta}=0.2$. In Fig.~(6.b), the same histogram is 
shown after
direct processes are transformed away. The Brody parameter in this case,
${\beta}=0.116$,
which is closer to a Poisson distribution (the Brody
distribution
becomes a Poisson distribution for $\beta=0$).

The distribution of nearest
neighbor eigenphase spacings has been computed for a energy independent
scattering matrix in \cite{kn:jung} for a very different physical 
system. They also
report close agreement with COE predictions for the chaotic region, 
although they do
not have to deal with direct processes. It is clear that direct 
processes can play an
important role in causing deviations from random matrix theory predictions for
scattering processes.

\section{Partial Delay Times}

In this section we compare the partial delay time distribution, computed
for the
deterministic scattering process, to values obtained  from random
matrix theory.  The partial delay times are given by the energy derivative of
eigenphases,
$\tau_{\alpha}=\hbar{d{\delta}_{\alpha}\over dE}$~\cite{kn:bohm}.
The average partial
delay time
density  for a scattering process governed by the scattering matrix,
${\bar
S}_{goe}$, has been computed by \cite{kn:fyo1,kn:fyo2}, using supersymmetry
techniques, and is given by,
\begin{equation}
{\rho}(\tau)=(1/N)\sum_{\alpha}{\langle}{\delta}(\tau
-\tau_{\alpha}){\rangle}_{goe}=
\frac{(1/2)^{N/2}}{\Gamma(N/2)}\frac{\exp(-1/2(\tau/{\langle}\tau{\rangle}))}{(\tau/{\langle}\tau{\rangle})^{N/2+2}}
\label{eq:td1}
\end{equation}
where ${\langle}\tau{\rangle}=1/N$.

Before showing the distribution obtained for the partial delay times, 
it is useful to
discuss energy correlations contained in the partial  delay time curves.
In Fig.~7, we show the auto-correlation function for the partial 
delay times obtained
in the 16 channel energy regime $256<E/E_1<289$ and averaged over
6 different ripple sizes, ${\rm a}=22\AA,~23\AA,~24\AA,~25\AA,~26\AA,~27\AA$. For each
partial delay time curve we obtain a auto-correlation function, and then we  average
over all 96 curves. 
We also show the GOE  prediction for the 16 channel case as well as the partial delay
time auto-correlation function for
the near integrable regime (${\rm a}=1\AA$). The GOE prediction is obtained after
performing the triple integration given in reference \cite{kn:leh}. The energy scale is
adjusted to correspond to the relevant scale for our data.  We also note the partial
delay time auto-correlation function is  the second derivative of the eigenphase
 auto-correlation function. Therefore the eigenphase auto-correlation function decays more slowly
than the partial time delay auto-correlation function.

In Fig.~(8.a), we show a histogram of the scaled partial delay times,
$\tau/{\langle}\tau{\rangle}$.
We again
consider  the energy
regime with 16 channels and vary the energy in the
interval, $256E_1<E<289E_1$. To obtain enough values  to build
good statistics,
we use 4 different ripple sizes, ${\rm a}=25\AA,~30\AA,~35\AA,~45\AA$.
For these ripple
amplitudes, the ripple cavity dynamics is chaotic.   We
used 100  energy points per specific ripple size, and therefore an 
energy increment
of $\Delta E=0.33E_1$. The average, delay time, ${\langle}\tau{\rangle}$, is 
obtained numerically
for each partial  delay time curve. Then the histograms for 64
scaled partial delay times are combined into one histogram by simply 
adding values in the corresponding bins.
  The solid line in Fig.~(8.a), is a plot of the RMT prediction, 
$N(\tau)={\eta}'{\rho}(\tau)$, where ${\eta}'$ is the area under the curve and 
${\langle}\tau{\rangle}=1/N$. The agreement
is not good because our data contains the effect of
direct scattering processes. In Fig.~(8.b) we show the partial delay time
density obtained
from the eigenphases, ${\theta}_{\alpha}$, which no longer contain the
effect of direct
scattering processes. The solid line is a plot of $N(\tau)$ with
${\langle}\tau{\rangle}=1/N$ \cite{kn:fyo1}. The agreement is
very good.  Finally in
Fig.~(8.c) we plot histogram of 4000  partial delay times obtained from
 a $16{\times}16$ ${\bar S}_{goe}$ by using different realizations.
Again, the agreement between the data and Eq. (\ref{eq:td1})  is very good.
    Thus, after the removal of the effects of direct
scattering processes
our deterministic scattering from the chaotic ripple cavity behaves very
much like the
RMT prediction.
(It is useful to note that in Ref. \cite{kn:akg1}, we compared the
{\it Wigner-Smith} delay
time distribution with numerically computed predictions of RMT. The
Wigner-Smith delay
time is defined,
${\tau}_{ws}={1\over
N}{\sum_{\alpha=1}^N}{\tau}_{\alpha}$.)

In Fig. 9, we show the delay time distributions
for the near integrable case, ${\rm a}=1\AA$.  We have used energy increments,
${\Delta}E=0.1E_1$ justified from Fig.~7.
  The delay time distribution for the near integrable case deviates significantly from the  random
matrix result and the results for chaotic cavity shown in Fig. (8).

Let us now return to the eigenphase curve in Fig. (3.b). We see that 
the curves for the near integrable case have a sequence of fairly abrupt large changes 
of phase. These are due to resonance structures that cause larger than average delays of 
the particle in the cavity. In Fig. (10), we plot the Wigner-Smith delay time (which is 
an average over all partial delay times) in the energy interval, 
$257.4{\leq}E/E_1{\leq}259.5$. This energy interval contains two of the large phase
changes in the eigenphase  curves in Fig. (3.b).
We see that each large phase change gives rise to a large peak in the 
delay time. The crosses in Fig. (10) give the energies of the cavity basis states, 
$\lambda_j$, in that energy interval. There appears to be one cavity state which lies at 
each resonance energy. In Fig. (10), we have also plotted the configuration space 
distribution of four of the cavity eigenstates, two at resonance and two off resonance.
In  Figs. (2.b)-(2.e), we have shown Husimi plots of the quantum Poincare surface of 
section for each of these four states. The two states giving rise  to the delay time
resonance  peaks lie in the dominant nonlinear resonance structures in the classical
phase space.  The quantum particle appears to tunnel into these dynamical resonance
structures,  and is delayed there for a considerable length of time.

\section{Conclusion}

We have analyzed the statistical properties of a scattering
process in a waveguide with a cavity which allows a range of 
dynamics, including
integrable, mixed, or chaotic. In this waveguide, direct processes 
also play an important
role.  The ``ripple" cavity that we use has the special feature that 
it allows us to form
a Hamiltonian matrix to describe the dynamics interior to the cavity. 
This, in turn,
allows us to use the reaction matrix approach to scattering for our 
deterministic
scattering process.  The reaction matrix approach is one of the most 
efficient methods
for obtaining the large amounts of data necessary to obtain good statistics.
Until now, mesh based models (like the boundary element method, finite 
element method, or
recursive Green's function method) were the main numerical methods to deal with
scattering problems. However, these methods use an energy dependent 
boundary condition
which makes it a formidable task to obtain solutions for very large 
numbers of energy
points. The reaction matrix approach allows us to circumvent this problem.
It is also useful to note that the reaction matrix approach has been used extensively to study
properties of the complex poles of the S-matrix. This is discussed in some detail in 
\cite{kn:jung,kkn:rotter1,kkn:rotter2,kkn:stockman2002}.

We have obtained a number of results. We find that, in the near 
integrable regime,
nonlinear resonances in the classical phase space give rise to large eigenphase
excursions and long delay times for quantum particles that can tunnel 
into these
dynamical structures.

We have focused much of our discussion on the energy regime in
which sixteen channels are open in the lead. We have been  
able to follow each
eigenphase of the S-matrix continuously as a function of energy. We 
have examined the
statistical properties of the scattering process by gathering data about each
eigenphase at  discrete energy intervals in the sixteen channel 
regime. This assumes a
kind of ``stationarity" as a function of energy, of the underlying
scattering process.   We have chosen the energy intervals so that our 
data points are
statistically independent.

We have shown, for the scattering system considered here,  that the affect of direct
processes on the eigenphase  curves can be
transformed away. We find that, for the case where the cavity 
dynamics is classically
chaotic, a partial time delay density histogram, formed  from all 
sixteen transformed
eigenphase curves, agrees to 96$\%$ confidence level with  a 
Brody distribution
with Brody parameter, ${\beta}=0.87$. 
Similar deviations
from the GOE prediction of ${\beta}=0.95$ have been seen in the 
nearest neighbor energy
eigenvalue spacing distributions of closed ripple billiards \cite{kn:li} and in that
case are  caused by bouncing
ball orbits. We expect the same mechanism is having an affect here.

\section {Acknowledgements}

The authors wish to thank the Welch Foundation, Grant
No.F-1051; NSF Grant INT-9602971;  and DOE contract No.DE-FG03-94ER14405
for partial support of this work. We also thank the University of
Texas at Austin High Performance Computing Center for use of their computer
facilities.

\pagebreak

\pagebreak

\newpage

\begin{figure}
\caption{The geometry of the two dimensional ballistic waveguide
used in our calculations; $a$ is the half-width of the  ripple,
$d$ is the width of the lead and the average width of the cavity. The ripple cavity
extends from $x=0$ to $x=L$}
\label{figure1}
\end{figure}

\begin{figure}
\caption{Surfaces of section for $L=511\AA$, $d=101\AA$, and ${\rm a}=1\AA$.
(a) A Poincare surface of section showing $p_x/p={\cos}(\alpha)$ 
versus $x$, each time
the particle hits the bottom wall. (b) Husimi plot of quantum surface 
of section (QSS)
for cavity eigenstate with ${\lambda}_j=257.1934E_1$. (c) QSS for 
cavity eigenstate with
${\lambda}_j=257.9339E_1$. (d) QSS for cavity eigenstate with
${\lambda}_j=258.6655E_1$. (e) QSS for cavity eigenstate with
${\lambda}_j=258.9072E_1$.}
\label{figure2}
\end{figure}

\begin{figure}
\caption{Eigenphases, ${\delta}_{\alpha}$ versus $E/E_1$ for the
energy interval,
$256{\leq}E/E_1{\leq}272$: (a)
${\rm a}=25\AA$,  (b) ${\rm a}=1\AA$.}
\label{figure3}
\end{figure}
\begin{figure}
\caption{ (a) Histogram of number of eigenphases, $N(\delta)$, versus $\delta$  for the
16 channel energy interval $256{\leq}E/E_1{\leq}289$, and for four
   different ripple sizes, ${\rm a}=25\AA,~30\AA,~35\AA,~45\AA$. The solid line is a
plot of the single channel Poisson kernel with
${\langle}S{\rangle}={\sum_{\alpha=1}^{64}}{\langle}{\rm
e}^{i{\delta}_{\alpha}}{\rangle}$ and normalized to the number of eigenphases.   (b)
Histogram of transformed eigenphases, ${\theta}_{\alpha}$, for all 64 eigenphase curves.
For all cases, $d=101\AA$ and  $L=511\AA$. A $\chi^2$ test result is also shown for both plots
 with 17 bins taken into account}
\label{figure4}
\end{figure}

\begin{figure}
\caption{ Histogram of number, $N({\sigma})$, of nearest neighbor scaled eigenphase
spacings,
${\sigma}$,  for the chaotic regime with
$d=101\AA$, $L=511\AA$. The average spacing, ${\langle}{\sigma}{\rangle}$ is obtained for
each eigenphase curve.  The histograms contain a total of $15{\times}67=1005$ data
points averaged over four different ripple sizes, ${\rm
a}=25\AA,~30\AA,~35\AA,~45\AA$. (We obtain a histogram for each of the four 
values of the ripple amplitude. We then add them and divide by four.)
 (a) Before direct processes are transformed away. The
thin solid line is the Brody distribution with $\beta=0.635$.
(b) After direct processes transformed away. The thin solid line
is the Brody distribution with $\beta=0.865$.A $\chi^2$ test result is also shown for both plots
 with 13 bins in it taken into account
}
\label{figure5}
\end{figure}
\begin{figure}
\caption{  Histogram of number, $N({\sigma})$, of nearest neighbor eigenphase spacings,
${\sigma}$, 
for the near integrable  regime with
$d=101\AA$, $L=511\AA$. The histograms contain a total of 1005 data
points averaged over
6 different ripple sizes, ${\rm a}=0.5\AA,~1\AA,~2\AA,~3\AA,~4\AA,~5\AA$.
(a) Before direct processes are transformed away. The thin solid line
is the Brody distribution with $\beta=0.2$.
(b) After direct processes are transformed away. The thin solid line
is the Brody distribution with $\beta=0.116$. A $\chi^2$ test result is also shown for both plots
 with 8 bins taken into account }
\label{figure6}
\end{figure}

\begin{figure}
\caption{The auto-correlation function of  time delays
in the 16 channel energy interval, $256<E/E_1<289$.
The thin line is  obtained numerically for the chaotic regime, using six
different ripple amplitudes, ${\rm a}=22\AA,~23\AA,~24\AA,~25\AA,~26\AA,~27\AA$, with
direct processes transformed out of the data. An  auto-correlation function is
obtained for each partial time delay curve and the average of those 96 auto-correlation
functions is shown. 
   The dotted-dashed line
is the GOE result for perfect coupling with 16 modes with the average density of
states chosen equal to $1.25$. The thick line represents the 
numerically obtained auto-correlation function in the region of mixed phase space for 
${\rm a}=0.5\AA,~1\AA,~2\AA,~3\AA,~5\AA$ }
\label{figure7}
\end{figure}

\begin{figure}
\caption{Histogram of number of scaled partial delay times, $N(\tau)$, versus
$\tau/{\langle}\tau{\rangle}$, for the 16 channel energy interval,
$256{\leq}E/E_1{\leq}289$, with $d=101\AA$ and $L=511\AA$. Data for
ripple amplitudes
${\rm a}=25\AA,~ 30\AA,~ 35\AA, ~45\AA$ is included in the histograms. 
Data points are taken at energy intervals, ${\Delta}E=0.33E_1$. 
 (a) Histogram of scaled partial delay times taken from 
eigenphase curves for
${\delta}_{\alpha}$. A scaling factor, ${\langle}\tau{\rangle}$, is obtained for each
eigenphase curve. (b) Histogram of scaled partial delay times taken from eigenphase
curves for the transformed eigenphases, ${\theta}_{\alpha}$. (c) Histogram of partial delay
times obtained from the $16{\times}16$ S-matrix, ${\bar S}_{goe}$ (includes 4000 data
points). A $\chi^2$ test result is also shown for the  plots (b) and (c)
 with 13 bins taken into account. 
}
\label{figure8}
\end{figure}
\begin{figure}
\caption{Histogram of partial delay times for the 16 channel energy interval,
$256{\leq}E/E_1{\leq}289$, with $d=101\AA$ and $L=511\AA$. Data for
ripple amplitudes, 
${\rm a}=0.5\AA,1\AA,2\AA,3\AA,5\AA$, is used to construct the histograms.
Data points are taken at energy spacings, ${\Delta}E=0.1E_1$. A total of
$400 X  16$ points is used.
 (a) Histogram of scaled partial delay time curves taken from eigenphase 
curves for ${\delta}_{\alpha}$. (b)
Histogram of scaled partial delay times taken from curves for transformed eigenphases, 
${\theta}_{\alpha}$.}
\label{figure9}
\end{figure}
\begin{figure}
\caption{Plot of Wigner-Smith delay time in the energy the interval
$257.4{\leq}E/E_1{\leq}259.5$  for  $d=101\AA$, $L=511\AA$, ${\rm a}=1\AA$. 
Crosses show values
of cavity basis state energies in this interval. Inserts show the 
spatial distribution
of four cavity basis states, two at resonance and two off resonance. }
\label{figure10}
\end{figure}
\end{document}